\pdfoutput=1
\documentclass[twocolumn]{aastex701}

\usepackage{amsmath}
\usepackage{graphicx}
\usepackage{booktabs}
\usepackage{placeins}
\usepackage{hyperref}
\hypersetup{colorlinks=true,citecolor=blue,urlcolor=blue,linkcolor=blue,breaklinks=true}

\newcommand{\code}[1]{\texttt{#1}}

\newcommand{\mesa}{\code{MESA}}

\newcommand{\ppmstar}{\code{PPMstar}}

\newcommand{\D}{{\mathrm d}}

\newcommand{\mach}{\ensuremath{{\mathcal Ma}}}

\newcommand{\brunt}{Brunt-V\"ais\"al\"a}

\let\tablenum\relax
\input{hydropaper-resources/macros/units}
\renewcommand{\erg}{\unitstyle{erg}} 
\input{hydropaper-resources/macros/vectors}
\newcommand{\isofont}[1]{{\mathrm{#1}}}
\newcommand{\isomass}[1]{{\ensuremath{\isofont{^{#1}}}}}
\newcommand{\isocharge}[1]{{\ensuremath{\isofont{_{#1}}}}}
\newcommand{\isotope}[3]{{\ensuremath{\isocharge{#1}\isomass{#2}\isofont{#3}}}}

\newcommand{\I}[2]{{\isotope{}{#1}{#2}}}

\newcommand{\Ep}[1]{{\ensuremath{10^{#1}}}}
\newcommand{\E}[1]{{\ensuremath{\powersep\Ep{#1}}}}

\newcommand{\powersep}{\times}

\definecolor{NavyBlue}{cmyk}{0.94,0.54,0,0}
\definecolor{BrickRed}{cmyk}{0,0.89,0.94,0.28}

\usepackage[normalem]{ulem} 

\graphicspath{{figures/}}

\submitjournal{ApJL}

\begin{document}

\title{The physics of deep-interior convective shell boundaries in a
  massive pre-supernova star: Internal gravity wave mixing at the $N^2$
  peak}

\shorttitle{Convective boundary mixing by internal gravity waves in deep-interior burning shells}
\shortauthors{Pathak et al.}

\author[orcid=0009-0005-9976-1882]{Praneet Pathak}
\affiliation{Astronomy Research Centre, Department of Physics \& Astronomy, University of Victoria, V8P 5C2, Victoria, Canada}
\email[show]{praneetpathak@uvic.ca}

\author[orcid=0000-0001-8087-9278]{Falk Herwig}
\affiliation{Astronomy Research Centre, Department of Physics \& Astronomy, University of Victoria, V8P 5C2, Victoria, Canada}
\email{fherwig@uvic.ca}

\author{Paul R. Woodward}
\affiliation{LCSE and Department of Astronomy, University of Minnesota, Minneapolis, MN 55455, USA}
\email{woodw001@umn.edu}

\author[orcid=0000-0002-1283-6636]{Joshua Issa}
\affiliation{Astronomy Research Centre, Department of Physics \& Astronomy, University of Victoria, V8P 5C2, Victoria, Canada}
\email{joshuaissa@uvic.ca}

\author[orcid=0000-0002-9632-1436]{Simon Blouin}
\affiliation{Astronomy Research Centre, Department of Physics \& Astronomy, University of Victoria, V8P 5C2, Victoria, Canada}
\email{sblouin@uvic.ca}

\begin{abstract}
The physics of convective boundary mixing involves concepts such as
penetration and overshooting and has been mostly characterized for
envelope and core convection. For the burning shells deep inside
massive stars in their final stages, the convective boundary remains
poorly understood. 1D stellar-evolution codes typically place a steep
entropy gradient and the corresponding narrow \brunt{} ($N^2$) peak at
a lower convective shell boundary. Based on high-resolution 3D
hydrodynamic simulations of a $25\usp\Msun$ star 11 days before core
collapse, we find that convection excites high-frequency internal
gravity waves at the $N^2$ peak representing the C-shell convection
bottom. These waves efficiently mix entropy and species and the
$N^2$ peak erodes by ${\sim}40\%$ within hours, correspondingly
reducing the entropy gradient. The wave-driven diffusion
coefficient is $D\approx 5\text{--}6\times
10^{9}\usp\power{\cm}{2}\usp\power{\second}{-1}$ at the $N^2$ peak in
numerically converged simulations. Our simulations show that the
dominant physics at this boundary is mixing due to internal gravity
waves. Extrapolated with the time-dependent $D$ coefficient from the
3D simulations, the entropy step across the peak decreases by 40--85\%
within half a day, thereby reducing the density, temperature and
entropy gradients that set the accretion-rate drop a stalled supernova
shock encounters. This physics is not included in present-day 1D
models and enters the structural parameters that predict
explodability. Flattening the gradients days before collapse favors an
O--C shell merger, which could reshape pre-collapse
nucleosynthesis. Without this physics 1D models carry too sharp
gradients that introduce previously unaccounted systematic errors into
the explodability surveys, yield tables and chemical-evolution studies
built on them.
\end{abstract}


\section{Introduction} \label{sec:intro}

The architecture of nuclear-burning shells in the final hours of a
massive star's life sets the pre-collapse structure that determines its
compactness, its nucleosynthetic yield, and whether and how it explodes
\citep{OConnor:2011,Couch:2013,Sukhbold:2014,Ertl:2016,Muller:2016PASA,Sukhbold:2016,Muller:2020LRCA}.
Near a zero-age main-sequence mass of $25\usp\Msun$ explodability
is highly non-monotonic in initial mass
\citep{Sukhbold:2014,Patton:2020}, so small changes in the late
convective shells can flip a model from successful explosion to direct
collapse. The thin radiative zones between these shells set the
entropy and composition gradients that determine whether shell mergers
occur and thereby
reshape the pre-collapse nucleosynthesis of intermediate and odd-$Z$
elements \citep{Ritter:2018,Roberti:2025}. In 1D stellar-evolution
models the pre-collapse structure depends sensitively on the convective
boundary mixing (CBM) prescription \citep{Davis:2019},
with reported variation in pre-collapse yields up to orders of magnitude
for the $p$-nuclei, the light odd-$Z$ isotopes including
$^{40}$K, and pre-explosive $^{44}$Ti that can rival the
supernova-explosion contribution, depending on the choice of
3D-informed inter-shell mixing \citep{Issa:2025b,Issa:2025,Issa:2026}.
The layer between the inner O-burning convective shell and the overlying
C-burning shell, with its narrow radiative zone and the steep entropy
step at its top, is thus consequential for pre-collapse
nucleosynthesis and explodability yet poorly constrained
\citep{Boccioli:2023}.

A recurring theme of the 3D stellar-hydrodynamics literature is that the
radial stratification produced by 1D codes, in which convection is
parameterized and internal-gravity-wave (IGW) feedback is not modeled from first principles, is generally not a stationary state of the 3D equations. The 3D system
then relaxes toward a different mean stratification
\citep{Anders:2022ApJL,anders:23,Herwig:23a,Lecoanet:2026}.
The same
picture has emerged
across advanced-burning boundaries and into
dynamical shell mergers
\citep{Cristini:2017,Mocak:2018,Cristini:2019,Andrassy:2020,Yadav:2020,andrassy:22,Rizzuti:22,Rizzuti:2023,Rizzuti:2024}.
Across that literature the mixing between the advanced shells is
attributed to entrainment at the convective boundaries.
The entrainment rate was measured in a 3D oxygen-shell model by
\citet{Meakin:2007}, who set it by the bulk Richardson number. In the
oxygen-neon shell merger of \citet{Yadav:2020} convective eddies
entrain neon and carbon into the oxygen shell. \citet{Meakin:2006}
simulated concurrent oxygen and carbon shells and found that the wave
motions excited at the convective boundaries reach the intervening
non-convective region, with consequences for compositional mixing
there.
\citet{Mao:2024} interpreted the large
entrainment rates measured for a $25\usp\Msun$ main-sequence model as the
3D response to a 1D initial stratification that is out of dynamic and
thermal equilibrium. IGW-driven chemical mixing in radiative zones has
been studied theoretically for decades
\citep{press:81,Lecoanet:2013hc,rogers:17,Edelmann:2019jh,Horst:2020ds,LeSaux:2022,LeSaux:2023,Vanon:2023,Morison:2024}.

CBM has been studied at the base of convective envelopes and at the
edge of convective cores \citep[reviewed for main-sequence stars
by][]{anders:23}. Below envelope and surface convection zones,
simulations describe it as overshooting
\citep{Freytag:vw,korre2021dynamics} and penetration
\citep{pratt2017extreme,Baraffe2021}. \citet{Anders:2022ApJ}
parameterized convective penetration and tested the theory in idealized
3D simulations. Applied to early-type main-sequence stars, that theory
gives enough mixing to explain the core masses inferred from
asteroseismology and eclipsing binaries, which suggests that most CBM at
their convective cores arises from penetration \citep{Jermyn:2022}. In
3D simulations of core convection, a nearly adiabatic layer develops
beyond the Schwarzschild boundary, a process called convective
penetration \citep{Anders:2022ApJ,Mao:2024}.
Core-overshooting parameters for 1D models have been derived from 2D
simulations \citep{Higl:2021,Baraffe:23} and from eclipsing binaries
\citep{Claret:2019ez}, and convective penetration has been implemented
in 1D stellar evolution \citep{Johnston:24}. CBM in the advanced burning
stages of massive stars is not well understood \citep{Davis:2019}. In 1D
models of these stages that treat CBM as an exponentially decaying
diffusion, the core masses at collapse, the compactness and the yields
depend strongly on its free parameter \citep{Davis:2019}.

\citet{Herwig:23a} showed, in 3D simulations of the convective core of a $25\usp\Msun$ main-sequence star, that the narrow $N^2$ peak just outside an adiabatic convective boundary is mixed by IGWs. Convective motions do not reach past this peak, where IGWs dominate the flow. In those simulations the boundary advanced into the stable envelope and eroded the mean molecular weight barrier as the 1D initial structure approached a thermal-dynamic equilibrium in 3D. IGW mixing set the shape of the migrating $N^2$ peak, but it was found not to be the dominant mixing physics in real main-sequence stars in thermal-dynamic equilibrium \citep{Mao:2024,Anders:2022ApJ}.

We present a suite of \ppmstar{} simulations of the
same $25\usp\Msun$ \mesa{} base state used by \citet{Jones:2017kc}, in
which both the O- and C-burning convective shells are driven
simultaneously at their nominal luminosities and the intervening
radiative layer is finely resolved. The study aims to identify the convective boundary mixing at the $N^2$ peak below the C-shell.
The
mixing is quantified by inverting the 1D diffusion equation for the
entropy variable to obtain a radial entropy-mixing diffusion
coefficient $D(r)$, following \citet{Herwig:23a}. A heating/no-heating control
run isolates the convection-driven component from thermal diffusion and
the relaxation of the initial stratification. An inert tracer fluid bounds the numerical
diffusion at the flow speeds the heated runs reach, and a resolution
series from $1280^{3}$ to $3072^{3}$ tests whether the inferred mixing is
physical or a numerical-diffusion artifact.
We \emph{do not} address extended secular integration to core collapse
(the runs span hours of simulated time), multiple ZAMS masses, rotation or
magnetic fields, and
the 3D opacity is a radius-only fit to the \mesa{}
profile.

\section{Numerical Setup and Simulation Suite}
\label{sec:setup}

The simulations use \ppmstar{}, an Eulerian piecewise-parabolic-method
(PPM) hydrodynamics solver on a Cartesian grid
\citep{Colella:84,Woodward:84,Porter_Woodward_1994,Woodward:15,Woodward2018},
previously applied to the same massive-star family
\citep{Jones:2017kc,Andrassy:2020,Herwig:23a,Andrassy:24,Mao:2024,Thompson:2023a,Pathak:2025}.
The 1D base state is \mesa{} model 28900 of the $25\usp\Msun$, $Z=0.02$
sequence
\citep{Jones:2017kc}, taken $11.4$ days before core collapse,
when an O-burning convective shell and an overlying C-burning convective
shell coexist.
That study simulated the O-shell alone with an ideal-gas equation of
state. Here we extend the domain outward to encompass the inter-shell
radiative layer and the base of the C-shell, and use an ideal-gas plus radiation-pressure equation of
state with radiation diffusion in the energy equation, as described by
\citet{Mao:2024}.
The radial domain runs from the bottom of the O-shell
($R_\mathrm{min}=4.5\usp\Mm$) to an outer boundary at
$R_\mathrm{max}=12\usp\Mm$ (small envelope) or $16\usp\Mm$ (large
envelope), both truncating the C-burning region. Two
inverted-parabola heating layers at the bases of the O- and C-shells
drive the two convection zones at their nominal luminosities
(Appendix~\ref{app:setup}, Fig.~\ref{fig:setup_inputs}b).
Two inert tracer-fluid (FV) Gaussians placed in the radiative layer measure $D$
there directly (Appendix~\ref{app:fv}), and a radius-only fit
approximates the \mesa{} opacity (Appendix~\ref{app:setup},
Eq.~\ref{eq:kappa_fit}). The run suite (Table~\ref{tab:runs}) spans
grids from $1280^3$ to $3072^3$ across both envelopes. O32 ($3072^3$,
small envelope) is the highest-resolution reference run, O31 is the
no-heating control, and O43 replaces the prescribed C-shell heating with
active $^{12}$C+$^{12}$C burning coupled to the flow following
\citet{Andrassy:2020} (Sect.~\ref{sec:burn}).

\begin{table}
\centering
\caption{Simulation suite. All runs use the same \mesa{} base
state and inner boundary $R_\mathrm{min}=4.5\usp\Mm$. The uniform grid
spacing is $\Delta x = 2R_\mathrm{max}/N$, and $t_\mathrm{end}$ is the
simulated time. O31 is the O29 setup with the O- and C-shell heating
switched off. O43 is the O28 setup with the C-shell driven by the
$^{12}$C+$^{12}$C burning network of Appendix~\ref{app:setup}.}
\label{tab:runs}
\begin{tabular}{lccccc}
\toprule
ID  & Grid     & $R_\mathrm{max}$ & $\Delta x$ & Env.   & $t_\mathrm{end}$ \\
    &          & (\Mm)            & (\Mm)      &        & (\hour)          \\
\midrule
O28 & $1536^3$ & 12 & 0.0156 & small & 1.76 \\
O29 & $1536^3$ & 16 & 0.0208 & large & 2.36 \\
O30 & $2048^3$ & 16 & 0.0156 & large & 1.54 \\
\shortstack[l]{O31\\ {\footnotesize (no heat)}} & $1536^3$ & 16 & 0.0208 & large & 2.54 \\
O32 & $3072^3$ & 12 & 0.0078 & small & 0.86 \\
O36 & $1280^3$ & 12 & 0.0188 & small & 1.81 \\
\shortstack[l]{O43\\ {\footnotesize (burn)}} & $1536^3$ & 12 & 0.0156 & small & 2.27 \\
\bottomrule
\end{tabular}
\end{table}

\section{Results}
\label{sec:results}

\subsection{Internal-Gravity-Wave Mixing at the $N^2$ Peak below the C-Shell}
\label{sec:igw_field}

\begin{figure*}[tb]
\centering
\includegraphics[width=\linewidth]{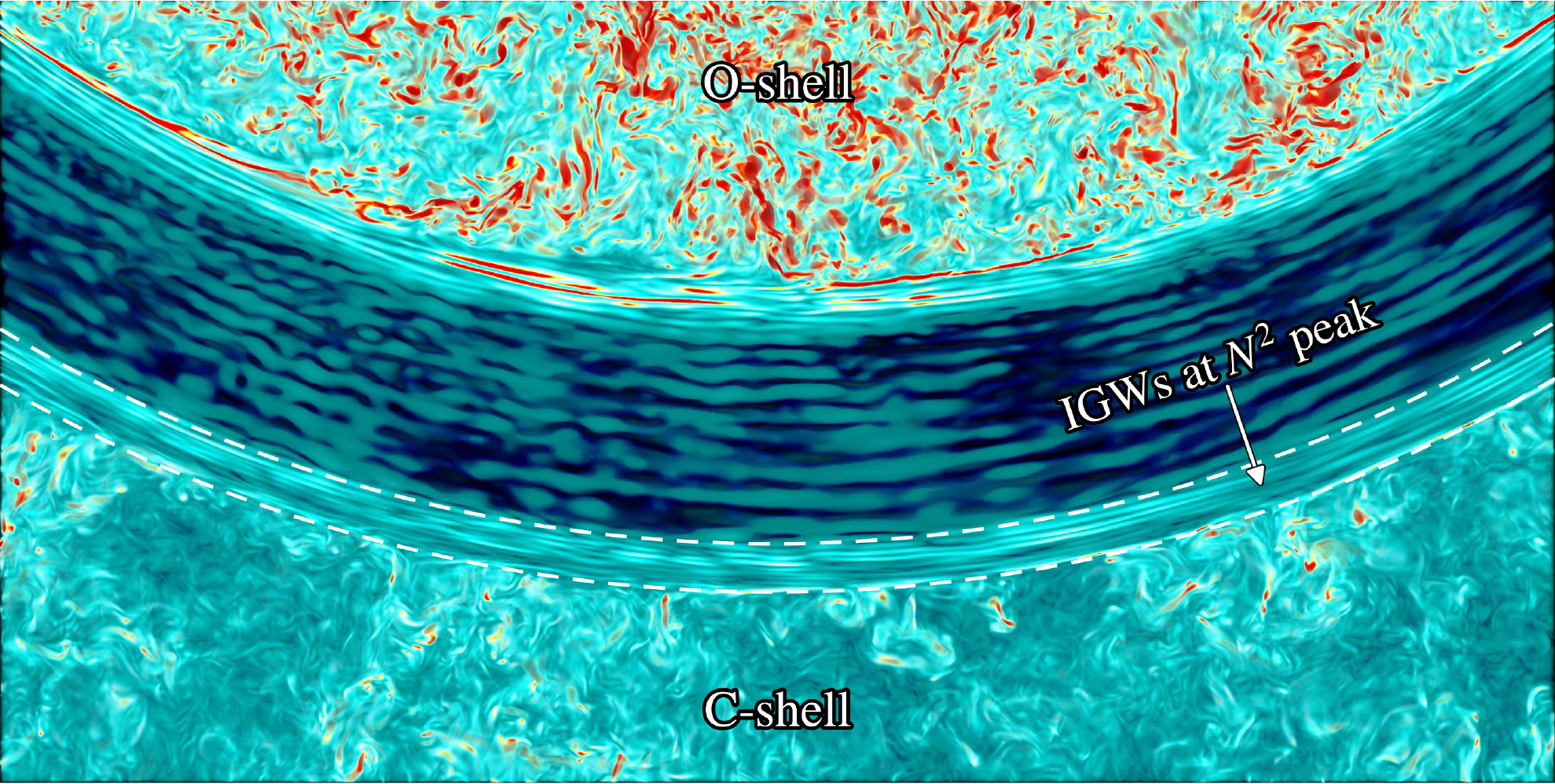}
\caption{Vorticity magnitude in a thin equatorial slice of run
O32 ($3072^3$) at $t=0.86\usp\hour$, zoomed on the inter-shell radiative
layer and colored from highest to lowest in red, yellow, light blue and
dark blue. Dashed white arcs at $r=9.3$ and $9.6\usp\Mm$ mark the $N^2$
peak just below the C-shell, the range over which $D>0$ in O32
(Fig.~\ref{fig:Dcombined}a).}
\label{fig:vort_zoom}
\end{figure*}

The two heated convective shells bound a narrow radiative layer in which
the \brunt{} frequency $N$ is of order $10^{5}\usp\mathrm{\mu Hz}$ and
peaks sharply at $\approx 6\times10^{5}\usp\mathrm{\mu Hz}$ at the base
of the C-shell (Appendix~\ref{app:setup}, Fig.~\ref{fig:base_state}b).
IGWs excited by O-shell convection cross this layer and reach that
$N^2$ peak.
A close-up of the vorticity magnitude in an equatorial slice of the
highest-resolution run O32 (Fig.~\ref{fig:vort_zoom}) shows O-shell and
C-shell convection as turbulent regions on either side of the radiative
layer. The layer carries a horizontally layered IGW field, and the
transition from convective turbulence to this IGW field is abrupt at each
boundary. Between the dashed arcs in Fig.~\ref{fig:vort_zoom}, the $N^2$
peak just below the C-shell carries strong IGWs with a finer horizontal
structure than the interior of the radiative layer.

\begin{figure}
\centering
\includegraphics[width=\linewidth]{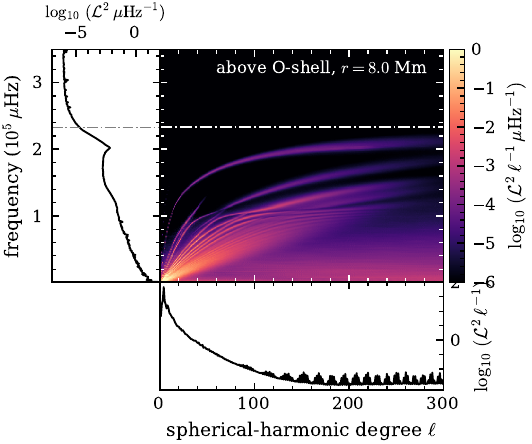}\\
\includegraphics[width=\linewidth]{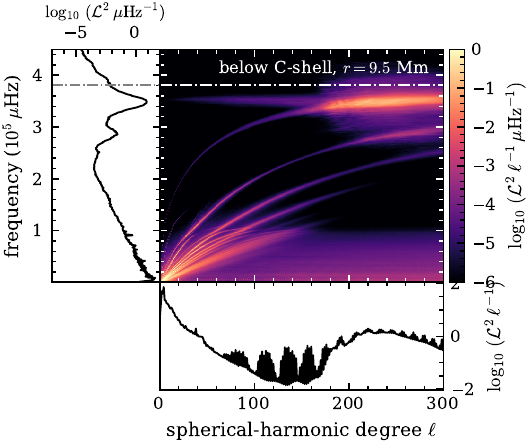}
\caption{$\ell$--$\nu$ diagrams of the unity-subtracted
relative luminosity $\mathcal{L}$ in O28 just above the O-shell
($r=8.01\usp\Mm$, top) and at the $N^2$ peak just below the C-shell
($r=9.50\usp\Mm$, bottom), over $t=1.27$--$1.74\usp\hour$. The
dot-dashed line marks the local \brunt{} frequency at
$t=1.74\usp\hour$. The side panels show the power integrated over $\ell$
(left) and over frequency (bottom).}
\label{fig:komega}
\end{figure}

Figure~\ref{fig:vort_zoom} shows three wave regimes. The interior
of the radiative layer carries a smooth large-scale pattern. The $N^2$
peaks at its two edges, one just above the O-shell and one just below the
C-shell, each carry a finer pattern of their own.
Their spectra differ, as the $\ell$--$\nu$ diagrams at $r=8.01$ and
$9.50\usp\Mm$ show (Fig.~\ref{fig:komega}). Appendix~\ref{app:wave} shows renderings of the full
equatorial slice (Fig.~\ref{fig:render_O32}) and a radius--frequency
propagation diagram of the layer (Fig.~\ref{fig:wave_prop}).

We characterize the spatial and temporal variability of the IGWs in the
radiative layer and at the $N^2$ peak with $\ell$--$\nu$ diagrams of the unity-subtracted
relative luminosity $\mathcal{L}=T^4/\langle T^4\rangle-1$, constructed
as in \citet{Thompson:2023a} and \citet{Pathak:2025} on spheres at fixed
radius for run O28 (Fig.~\ref{fig:komega}). Here $\ell$ is the
spherical-harmonic angular degree of the horizontal pattern, $\nu$ is the
temporal frequency, and $\langle T^4\rangle$ is the spherical average of
$T^4$ at that radius, which removes the global luminosity
trend. Just above the O-shell at
$r=8.01\usp\Mm$ the diagram is a clean set of dispersion ridges below
the local \brunt{} frequency, the standard IGW signature
\citep{Herwig:23a,Mao:2024,Thompson:2023a,Pathak:2025}.
At the $N^2$ peak just below the C-shell, $r=9.50\usp\Mm$, the same
low-$\ell$ ridges are present, but a pronounced concentration of power
appears at high angular degree, $\ell\approx 250$, close to the
\brunt{} frequency. This
corresponds to the horizontal scale of the thin radial shell in which
$N$ peaks (Fig.~\ref{fig:vort_zoom}), so the modes are confined to the narrow $N^2$ peak
immediately below the C-shell.

\subsection{Wave-Induced Rearrangement and a Heating/No-Heating Control}
\label{sec:rearrangement}

The IGW field reshapes the 1D-derived mean stratification. The 1D base state has steep steps in temperature,
entropy and density at the base of the C-shell, where $N$ peaks. Over
the simulated time these steps soften and the sharp $N$ peak, a
sensitive diagnostic since it depends on the entropy gradient, erodes
(Fig.~\ref{fig:profiles}, with the large-envelope runs and the
no-heating control in Appendix~\ref{app:control},
Fig.~\ref{fig:profiles_env}). At $t=0.86\usp\hour$ (the end of O32) and $t=1.2\usp\hour$ the
heated runs behave alike.
The $N$ peak falls from $N\approx5.8\times10^{5}\usp\mathrm{\mu Hz}$ at
$t=0$ to ${\approx}4\times10^{5}$ by $0.86\usp\hour$ and
${\approx}3.8\times10^{5}$ by $1.2\usp\hour$, a ${\sim}35\%$ reduction
that deepens toward ${\sim}40\%$ in the longer runs. The highest-resolution
run O32 is the least eroded at fixed time.

\begin{figure}[tb]
\centering
\includegraphics[width=\linewidth]{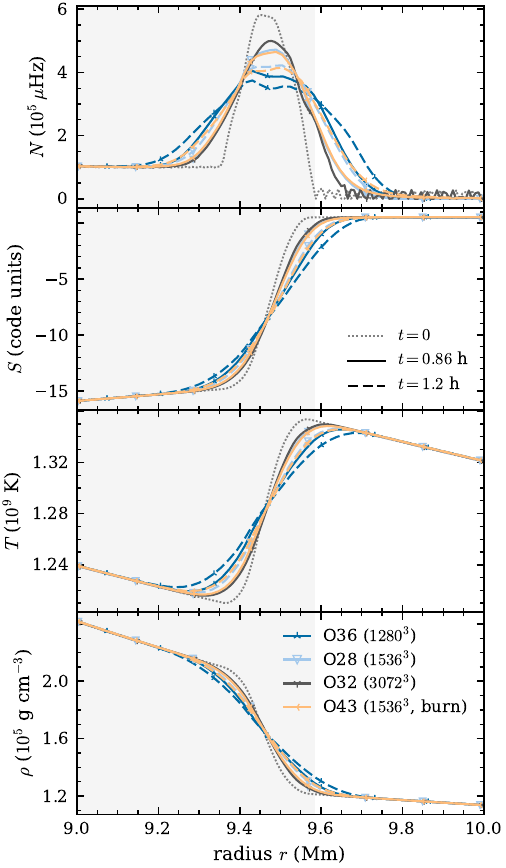}
\caption{Radial profiles of the \brunt{} frequency $N$ (top) and
the entropy $S$, temperature $T$ and density $\rho$ (bottom) across the
$N^2$ peak below the C-shell, for the small-envelope resolution sequence
O36, O28 and O32 and the active C-burning run O43. Each run (color, marker) is time-averaged over a
short window and shown at $t=0.86\usp\hour$ (solid) and $1.2\usp\hour$ (dashed), with the shared $t=0$ base state (grey
dotted). The grey band
marks the inter-shell radiative layer, which ends at $R\approx9.6\usp\Mm$.}
\label{fig:profiles}
\end{figure}

The same rearrangement could also be produced by thermal
diffusion or by the relaxation of the initial stratification. We ran an identical
experiment with the O- and C-shell volume heating switched off
(O31). The Mach numbers then drop by an order of magnitude
or more throughout the domain
(Appendix~\ref{app:setup}, Fig.~\ref{fig:velocity}a;
Appendix~\ref{app:control}, Fig.~\ref{fig:render_control}). The $N$ peak
and the temperature and entropy steps then stay close to their $t=0$
values. O31's peak falls only to ${\approx}5\times10^{5}\usp\mathrm{\mu Hz}$
by $1.2\usp\hour$ versus ${\approx}3.8\times10^{5}$ in the heated runs
(Appendix~\ref{app:control}, Fig.~\ref{fig:profiles_env}). The convection-driven IGW field drives the
rearrangement. The residual erosion in O31 bounds the
relaxation of the initial stratification and the thermal diffusion
together, both of which act whether or not the heating is on. The
convective spin-up is excluded instead by the inversion windows, which
begin at the velocity plateau (Appendix~\ref{app:dmethod}). Numerical
diffusion is bounded separately (\S\ref{sec:convergence}).

\subsection{The Mixing Diffusion Coefficient at the $N^2$ Peak}
\label{sec:Dcoeff}

\begin{figure*}
\centering
\includegraphics[width=\textwidth]{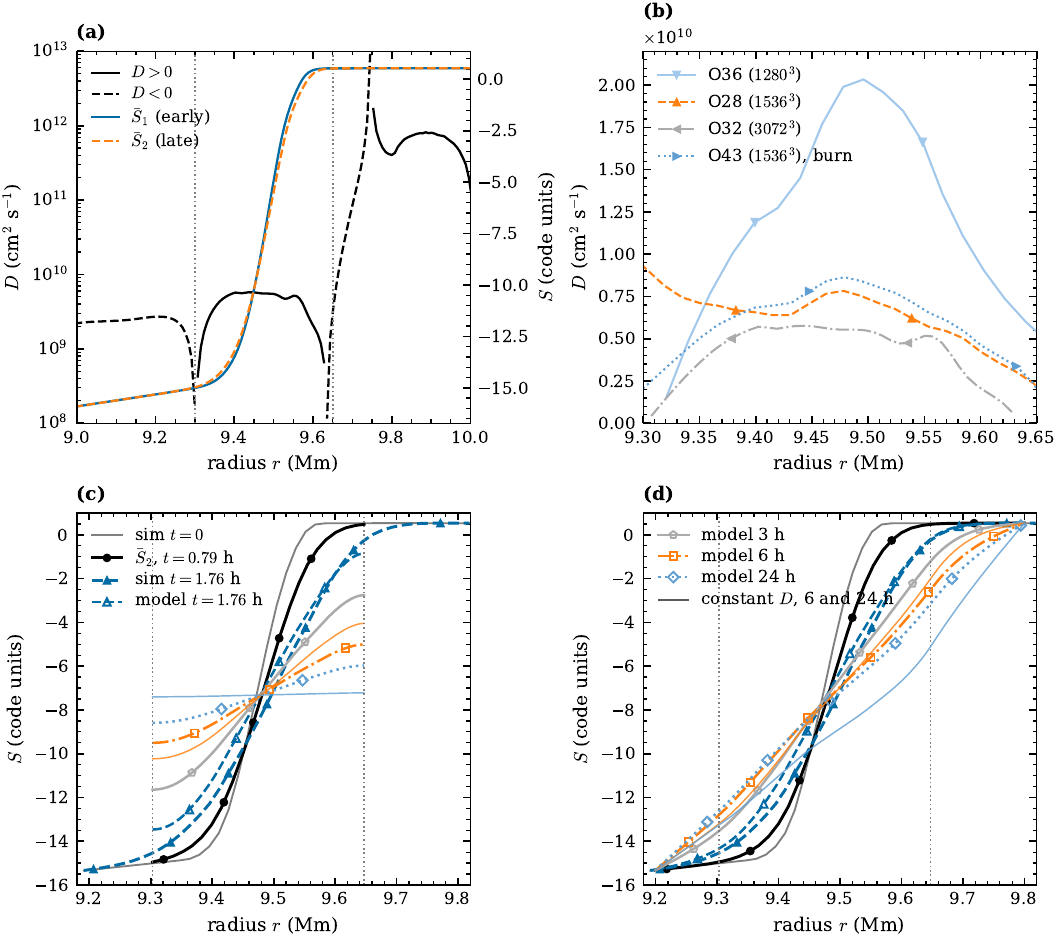}
\caption{The entropy-mixing diffusion coefficient at the $N^2$ peak. (a)
Inversion for O32 ($3072^3$). Entropy $S$ averaged over an early
($\bar S_1$, solid) and a late ($\bar S_2$, dashed) time window (right
axis), and the inferred $D(r)$ (left axis, solid where $D>0$ and dashed
where $D<0$). Dotted vertical lines mark the inversion
window $R\in[9.30,9.65]\usp\Mm$. (b) $D(r)$ for the small-envelope resolution sequence O36
($1280^3$), O28 ($1536^3$) and O32 ($3072^3$) and the active C-burning
run O43 ($1536^3$).
(c) Entropy of O28 evolved forward from $\bar S_2$
($t_2=0.79\usp\hour$) with its own $D(r,t)$, continued linearly to zero
at $11.9\usp\hour$ (Appendix~\ref{app:forward}), and with no entropy flux
through the edges of the window where $D>0$ (dotted). Filled markers are the simulation ($t_2$ and $1.76\usp\hour$) and
open markers the model ($1.76$, $3$, $6$ and $24\usp\hour$, unchanged
after $12\usp\hour$). Thin lines are the model with $D(r)$ held
constant, at $6$ and $24\usp\hour$. The grey line is the $t=0$ profile. (d) The same with the entropy held at its
$t_2$ value at $9.2$ and $9.8\usp\Mm$, and with $D$ continued at its
window-edge value outside the window. The two panels share one legend.}
\label{fig:Dcombined}
\end{figure*}

We quantify the relaxation as a radial diffusion coefficient $D(r)$,
obtained by inverting the 1D diffusion equation for the entropy variable
$S$ from its radial profile over $R\in[9.30,9.65]\usp\Mm$, around
the \brunt{} peak, following \citet{Herwig:23a}. Two independent
time windows, $W_1$ (early) and $W_2$ (late), set from the plateau of the
volume-averaged velocity (Fig.~\ref{fig:velocity}b;
Appendix~\ref{app:dmethod}), yield consistent profiles
(Fig.~\ref{fig:Dcombined}a). The overbar in $\bar S_1$ and $\bar S_2$
denotes the average of $S$ over the corresponding window. For the
highest-resolution run (O32, $3072^3$, $\Delta x\approx 0.008\usp\Mm$)
the peak value is
$D\approx 5\text{--}6\times 10^{9}\usp\power{\cm}{2}\usp\power{\second}{-1}$.
The large-envelope run O30 ($2048^3$), with the grid spacing
$\Delta x$ of the small-envelope O28, gives a peak
$D\approx 7\times 10^{9}$, close to the ${\approx}8\times 10^{9}$ of O28
(Appendix~\ref{app:dmethod}, Fig.~\ref{fig:D_envelope}).

\subsection{Is the Mixing Physical?}
\label{sec:convergence}

The \brunt{} peak below the C-shell is resolved by only a handful of
radial cells even at $3072^3$, so the inferred mixing could in principle
be numerical diffusion. A physical
effect superimposed on a numerical-diffusion floor would make $D$
\emph{decrease} with refinement toward an asymptote with shrinking
decrements, as in \ppmstar{} entrainment studies \citep{Woodward:15}.
A pure numerical artifact would keep $D$ flat or rising under
refinement. Along the small-envelope sequence
from O36 ($1280^3$) to O28 ($1536^3$) to O32 ($3072^3$) the peak $D$
\emph{decreases}, from $\approx 2.0\times10^{10}$ through
$7.8\times10^{9}$ to
$5.8\times10^{9}\usp\power{\cm}{2}\usp\power{\second}{-1}$, with the
decrement from O36 to O28 ($\approx 1.2\times10^{10}$) much larger
than that from O28 to O32 ($\approx 2.0\times10^{9}$)
(Fig.~\ref{fig:Dcombined}b). The
result is insensitive to the averaging-window width (${<}1\%$ for O32;
Appendix~\ref{app:dmethod}, Fig.~\ref{fig:Dwindow}).
Numerical diffusion acts only where the flow carries material
across the grid, so it is constrained by the resolution series above and
by the inert tracer fluid (FV), which the same flow advects in the heated
runs. The tracer erodes at
$D\approx(3\text{--}7)\times10^{7}\usp\power{\cm}{2}\usp\power{\second}{-1}$
at its two radii, two orders of magnitude below the value at the $N^2$ peak,
and falls with resolution as the $S$-based coefficient does
(Appendix~\ref{app:fv}, Fig.~\ref{fig:fvhhe}). We therefore adopt
$D\approx 5\text{--}6\times10^{9}\usp\power{\cm}{2}\usp\power{\second}{-1}$
as the converged, physical entropy-mixing rate at the $N^2$ peak.

\subsection{Active Carbon Burning}
\label{sec:burn}

Run O43 couples the $^{12}$C+$^{12}$C reaction to the
flow (Appendix~\ref{app:setup}, Eq.~\ref{eq:eps_nuc}) on the O28 grid
and envelope ($1536^3$, $R_\mathrm{max}=12\usp\Mm$), so the two runs
differ only in how the C-shell is driven. At $t=0.86$ and $1.2\usp\hour$ the O43 $N$ peak is
$4.65$ and $4.16\times10^{5}\usp\mathrm{\mu Hz}$ against $4.73$ and
$4.22\times10^{5}$ for O28, and its entropy, temperature and density
steps lie on those of O28 (Fig.~\ref{fig:profiles}). The entropy
inversion over the common time window gives a peak
$D\approx8.6\times10^{9}\usp\power{\cm}{2}\usp\power{\second}{-1}$,
about 10\% above O28's $7.8\times10^{9}$ and on the same downward trend
with resolution (Fig.~\ref{fig:Dcombined}b). The IGW mixing at the $N^2$
peak is therefore nearly the same whether C burning is approximated as a
constant volume heating or included in the simulation as a temperature-
and therefore time-dependent reaction source.

\subsection{Extrapolation of $S$ and the $N^2$ Peak}
\label{sec:fate}

To estimate the impact of the IGW boundary mixing over the
evolutionary time scale of the O-shell convection, we evolve $\bar S_2$
of O28 (\S\ref{sec:Dcoeff}) forward in time by solving the 1D diffusion
equation with a $D(r,t)$ taken from the run's own record
(Appendix~\ref{app:forward}). Sliding the two inversion windows through
O28's $1.76\usp\hour$ measures $D(r)$ at eleven times. It rises by a
factor $1.7$ from the common interval of Fig.~\ref{fig:Dcombined}b to
$1.2\usp\hour$, as the gradient it acts on flattens, and declines after.
The decline over the last four windows, $9.5\%$ per hour, continued
linearly brings $D$ to zero at $11.9\usp\hour$, after which the $N^2$ peak no
longer mixes (Fig.~\ref{fig:D_extrap}). Two edge treatments bound the
response of the adjacent convection zones. One passes no entropy flux
through the edges of the window in which $D>0$
($9.30$--$9.65\usp\Mm$), the other holds the entropy at $9.2$ and
$9.8\usp\Mm$, where the simulated profile does not change over the
run. Started from $\bar S_2$ at $t_2=0.79\usp\hour$, both
reproduce the simulated profile at $1.76\usp\hour$ to an rms of $0.7$
units of the $15.4$-unit entropy step across the window, and the second
also reproduces the entropy at the window edges, which the first
exceeds (Fig.~\ref{fig:Dcombined}c,d). By the time the mixing stops
the step across the window has shrunk to $2.6$ units with zero-flux
edges and to $9.6$ units with the reservoirs held, by $84\%$ and $40\%$,
and stays there. With $D$ held constant at its inverted value the
mixing continues instead. The zero-flux window flattens completely
within two days, and the held reservoirs leave a linear ramp of $7.5$
units.

\FloatBarrier
\section{Discussion and Conclusions}
\label{sec:discussion}

The high-resolution \ppmstar{} simulations show that the 1D-derived stratification at the $N^2$ peak below the C-burning shell of a
$25\usp\Msun$ pre-supernova model is not a stationary state of the 3D equations. Convection in the O-shell excites IGWs at the narrow $N^2$ peak just below the C-shell, with a
strong high-frequency and high-angular-degree ($\ell\approx 250$)
feature (Fig.~\ref{fig:komega}, bottom). The IGW field mixes entropy
in the $N^2$ peak. The steep entropy step at the base of
the C-shell softens and the $N^2$ peak erodes by ${\sim}40\%$ within
hours of simulated time. The relaxation is quantified by an
entropy-mixing diffusion coefficient, converged at
$D\approx 5\text{--}6\times 10^{9}\usp\power{\cm}{2}\usp\power{\second}{-1}$
in the highest-resolution run. Its downward convergence with refinement
identifies it as physical. The heating/no-heating control confirms that
the convection-excited IGW field drives the mixing. The convective
boundary mixing at the base of the C-shell is therefore carried by IGWs
rather than by overshoot, penetration or entrainment. With the prescribed C-shell
heating replaced by the $^{12}$C+$^{12}$C burning network (O43), $D(r)$ is
similar to that of the volume-heated run O28, its peak lying about 10\%
above O28's.

The finding extends the literature on 1D-to-3D stratification
relaxation at convective boundaries and in shell mergers
\citep{Meakin:2007,Cristini:2017,Mocak:2018,Andrassy:2020,Yadav:2020,Anders:2022ApJL,andrassy:22,Rizzuti:22,Rizzuti:2023,Andrassy:24,Rizzuti:2024},
to relaxation carried by IGWs at the convective boundary itself,
in the narrow $N^2$ peak at the base of the C-shell.

The reduced peak \brunt{} frequency at the base of the C-shell has
implications for shell mergers. 3D simulations
show that merging proceeds by entrainment that erodes the
stable layers separating the convective shells \citep{Rizzuti:2024}, and
the $N^2$ peak, or equivalently the steep entropy gradient, is the
barrier that has to be removed. IGW-driven mixing at the $N^2$ peak erodes that barrier in place, without the convective boundary having to advance by entrainment. On the measured $D(r,t)$ the entropy step across the $N^2$
peak shrinks by $40$--$85\%$ within half a day, before the mixing
itself dies away (\S\ref{sec:fate}, Fig.~\ref{fig:Dcombined}c,d). In the 1D
model the outer boundary of the O-shell and the base of the C-shell remain
at nearly the same radius for at least the next 3--4 days
\citep[Figs.~2 and 5 of][]{Jones:2017kc}, and core collapse is $11.4$
days away. The wave-driven mixing therefore flattens the $N^2$ peak below
the C-shell long before the 1D model moves either convective boundary,
and so acts in favor of a merger. How much mixing
occurs then sets the nucleosynthesis. \citet{Issa:2025b} find that the $^{40}$K yield of
an O-C shell merger varies by more than three orders of magnitude with
the assumed 3D mixing. That spread propagates to the galactic inventory
of the light odd-$Z$ elements P, Cl, K and Sc, which chemical-evolution
models underproduce relative to observations. As a radiogenic heat source,
$^{40}$K carries this spread into the thermal evolution of rocky
exoplanets. The same sensitivity appears in $^{44}$Ti.
\citet{Issa:2026} find a pre-explosive yield from an O-C shell merger
that can exceed the explosive production and match the observed values,
spread over nearly five orders of magnitude across 3D mixing scenarios.

The steep density drop across an
advanced-shell interface produces a sharp decline in the accretion rate
as it passes through the stalled shock, and that decline can help revive
the shock \citep{OConnor:2011}. The mixing measured here smooths that
drop before collapse. The critical neutrino luminosity, the luminosity
the proto-neutron star must supply for the stalled shock to run away,
scales as $L_\mathrm{crit}\propto\dot{M}^{0.79}$ in the runaway-bubble
model of \citet{Muller:2015}, so a factor-two drop in $\dot{M}$ at
interface passage lowers $L_\mathrm{crit}$ by a factor of ${\approx}1.7$. \citet{Boccioli:2023}
make the jump itself the criterion, reproducing the outcome of
${\sim}1300$ 1D models with neutrino-driven turbulence in more than
90\% of cases, so the coefficient measured here sets the rate at which that
jump is degraded. The $M_4$--$\mu_4$ criterion of
\citet{Ertl:2016} is likewise evaluated at an entropy step of the kind
eroded in these simulations. $M_4$ is defined at $s=4$, which they
identify as the base of the oxygen shell, and the companion parameter
$\mu_4=\mathrm{d}(m/\Msun)/\mathrm{d}(r/1000\usp\mathrm{km})$ is the
normalized mass derivative there, which they link to the mass-infall
rate. Smoothing the step therefore acts against the
revival trigger. Compactness has been argued to work the other way, a decrease favoring
explosion. \citet{Burrows:2025} obtain explosions at both low and high
compactness and reject compactness as an explodability condition. Because $\mu_4$ is evaluated as a finite
difference over $0.3\usp\Msun$, a smoothing width small compared with
that leaves the 1D criteria unchanged. The interface the shock
encounters is softened even so.

1D pre-supernova models do not model this IGW feedback from first principles and therefore carry systematically too-sharp $N^2$ peaks at these boundaries. The mixing measured here is the convective boundary mixing of these shells. It is the mechanism \citet{Herwig:23a} identified at the convective core boundary of a main-sequence star, now operating at the boundary of a deep-interior burning shell. The calibration effort now aimed at overshoot, penetration and entrainment \citep{Davis:2019,Kaiser:2020,Higl:2021,anders:23,Andrassy:24} should instead be built on the physics of IGW mixing.

Several limitations apply. The runs span hours of simulated time,
much shorter than the days remaining to core collapse. The opacity is
a radius-only fit. The O-shell heating is static in all runs, and the C-shell heating in
all but O43, which couples one $^{12}$C+$^{12}$C reaction. A single \mesa{} base state at one
ZAMS mass is used. Rotation and magnetic fields are neglected, the
latter shown to suppress O-shell entrainment by ${\sim}20\%$
\citep{Leidi:2023}. Only two envelope sizes
($R_\mathrm{max}=12$ and $16\usp\Mm$) are tested. The forward extrapolation of \S\ref{sec:fate} continues the
measured decline of $D$ linearly beyond the end of the run. The
response of the two convection zones is not modeled, and two edge
treatments stand in for it.

\section*{Data and Code Availability}

The simulation data underlying this work are publicly available from the
Federated Research Data Repository \citep[FRDR;][]{FRDR}, operated by the
Digital Research Alliance of Canada, at
\dataset[doi:10.20383/103.01782]{https://doi.org/10.20383/103.01782}
\citep{Pathak:2026data} under a CC~BY~4.0 license. The deposit contains, for each of the runs of
Table~\ref{tab:runs}, the spherically averaged radial profiles of every
dump of the run and the briquette-averaged 3D data cubes of its final
dumps (the last 50, except 30 for O30 and 10 for O32), together with the
1D initial stratification each run was started from.

The release follows the FAIR principles for scientific data management
and stewardship \citep{Wilkinson:2016}. Reader scripts
that use the open-source \code{ppmpy} package
(\url{https://github.com/PPMstar/PyPPM}) are included, pinned to the
commit they were validated against. \ppmstar{} itself is proprietary and
is not part of the release.

Movies of the simulations are available at \url{https://www.ppmstar.org/}.

\begin{acknowledgments}
P.~R.~Woodward acknowledges funding from NSF CDS\&E awards 1814181 and
2309101. The production simulations reported here were run on the
Frontera system at the Texas Advanced Computing Center, University of
Texas at Austin, supported by the National Science Foundation, and on
the Trillium system operated by SciNet at the University of Toronto
under the Digital Research Alliance of Canada. Volume renderings were generated with scripts
generously shared by Ted Wetherbee, to whom we are grateful. Analysis
was performed within the Astrohub virtual research environment
(\url{https://astrohub.uvic.ca/}), running on the Digital Research
Alliance of Canada Arbutus cloud at the University of Victoria, and on
the SciNet Open OnDemand portal
(\url{https://ondemand.scinet.utoronto.ca/}). Large language models
were used to refine wording at the sentence level and to assist with
coding.
\end{acknowledgments}

\software{\mesa{} \citep{Paxton:2011,Paxton:2013,Paxton:2015},
\ppmstar{} \citep{Colella:84,Woodward:84,Porter_Woodward_1994,Woodward:15,Woodward2018},
\code{ppmpy} (\url{https://github.com/PPMstar/PyPPM}),
NumPy \citep{Harris:2020},
Matplotlib \citep{Hunter:2007}}

\appendix
\makeatletter\@addtoreset{equation}{section}\makeatother

\section{Numerical Setup and Base State}
\label{app:setup}

\begin{figure*}[tb]
\centering
\begin{minipage}[t]{0.49\textwidth}\centering (a)\par\includegraphics[width=\linewidth]{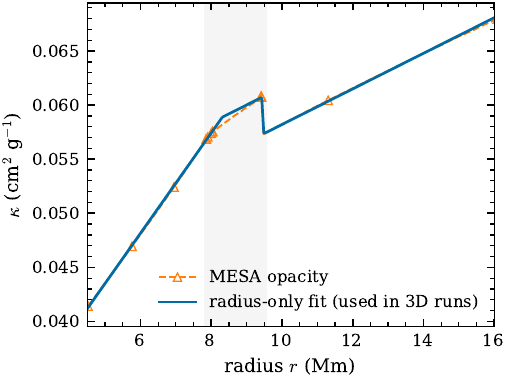}\end{minipage}\hfill
\begin{minipage}[t]{0.49\textwidth}\centering (b)\par\includegraphics[width=\linewidth]{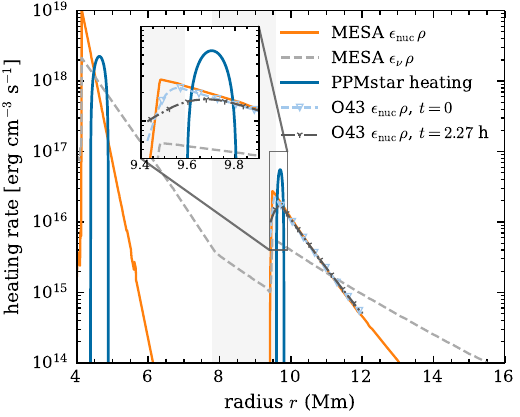}\end{minipage}
\caption{Prescribed radial inputs of the 3D runs. (a) Radius-only
opacity fit (Eq.~\ref{eq:kappa_fit}; solid line) and the \mesa{} opacity
it approximates (markers). The fit coefficients are
$a = 3.1424919\times 10^{-3}$,
$b = -1.4965588\times 10^{-3}$,
$c = 8.3146335$,
$d = 1.1287049\times 10^{3}$,
$e = -1.7269081\times 10^{-3}$,
$f = 6.8275601\times 10^{1}$,
$g = 9.4562937$,
$h = 3.1033979\times 10^{-2}$. (b) Nuclear energy-generation rate per
unit volume $\epsilon_{\mathrm{nuc}}\rho$ and neutrino energy-loss rate
$\epsilon_{\nu}\rho$ (dashed) of the \mesa{} base state, with the volume
heating rate deposited by \ppmstar{} in run O32 (log scale, in
\erg\usp\power{\cm}{-3}\usp\power{\second}{-1}). The two
$\epsilon_{\mathrm{nuc}}\rho$ peaks mark the O- and C-burning shells, and
\ppmstar{} drives the convection zones with inverted-parabola heating
layers at the base of each shell ($r\in[4.38,4.88]$ and
$[9.59,9.81]\usp\Mm$), each normalized to the nominal burning luminosity
($2.0\times10^{44}$ and $9.4\times10^{42}\usp\ergspersecond$). The
$^{12}$C+$^{12}$C burning rate of O43 (Eq.~\ref{eq:eps_nuc}, multiplied by
$\rho$) is shown at $t=0$ and at its last time, $t=2.27\usp\hour$. The
inset magnifies the base of the C-shell, $r=9.4\text{--}9.9\usp\Mm$ over
$4\times10^{15}\text{--}10^{17}\usp\erg\usp\power{\cm}{-3}\usp\power{\second}{-1}$,
where the prescribed heating layer, the \mesa{} profile and the O43
burning rate overlap. Its curves are those of the main panel and its
$y$ axis is unlabeled. Grey bands
mark the inter-shell radiative layer, as in Fig.~\ref{fig:profiles}.}
\label{fig:setup_inputs}
\end{figure*}

\begin{figure*}[tb]
\centering
\begin{minipage}[t]{0.49\textwidth}\centering (a)\par\includegraphics[width=\linewidth]{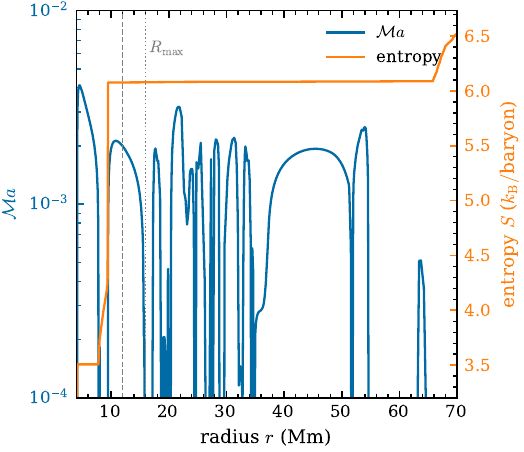}\end{minipage}\hfill
\begin{minipage}[t]{0.49\textwidth}\centering (b)\par\includegraphics[width=\linewidth]{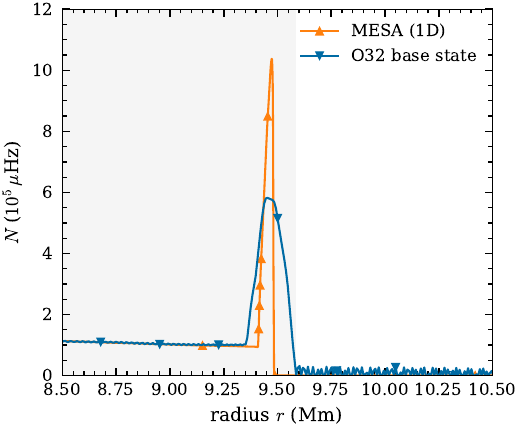}\end{minipage}
\caption{Stratification of the \mesa{} model 28900 used in this study.
(a) Convective Mach number $\mach$ (left axis) and entropy $S$ (right
axis) versus radius. The C-burning convective region (flat-entropy
plateau above the inter-shell radiative layer) extends out to
${\sim}65\usp\Mm$, far beyond the simulated outer boundaries
($R_\mathrm{max}=12$ and $16\usp\Mm$; vertical dashed and dotted lines).
(b) Initial \brunt{} frequency $N$ across the base of the C-shell, for
the 1D \mesa{} model (Ledoux $N^2$ including the gas/radiation
thermodynamic factors) and for the O32 base state loaded into the 3D run at $t=0$.
The grey band in (b) marks the inter-shell radiative layer, as
in Fig.~\ref{fig:profiles}.}
\label{fig:base_state}
\end{figure*}

\begin{figure*}[tb]
\centering
\begin{minipage}[t]{0.49\textwidth}\centering (a)\par\includegraphics[width=\linewidth]{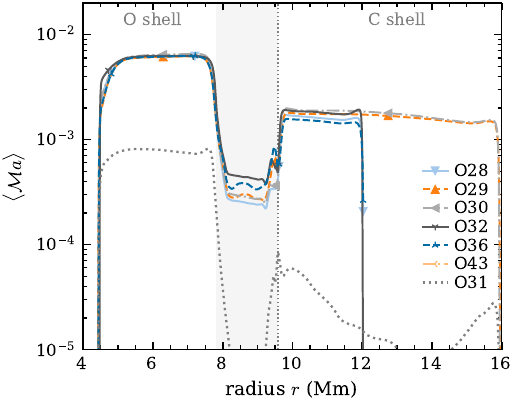}\end{minipage}\hfill
\begin{minipage}[t]{0.49\textwidth}\centering (b)\par\includegraphics[width=\linewidth]{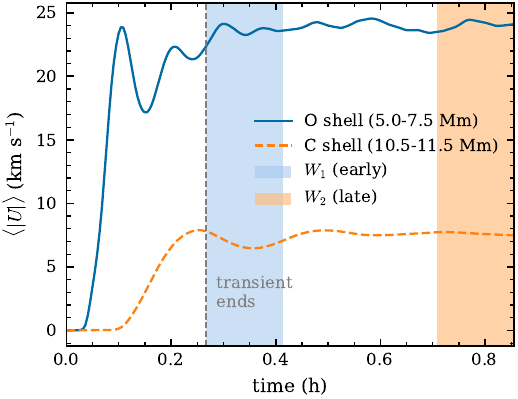}\end{minipage}
\caption{Convective velocity diagnostics. (a) Mach-number
profiles for the runs of Table~\ref{tab:runs}, each averaged over $t\in[0.30,0.85]\usp\hour$ (mean
$t=0.58\usp\hour$). The two convective shells appear as high-Mach plateaus
on either side of the low-Mach inter-shell radiative layer, and the dotted vertical line
marks the base of the C-shell. The
no-heating control O31 is the dotted profile. (b) Volume-averaged convective velocity
$\langle|U|\rangle$ versus time for O32, in the O-shell band
($5\text{--}7.5\usp\Mm$) and inside the C-shell
($10.5\text{--}11.5\usp\Mm$). After the initial transient the O-shell
velocity settles onto a plateau, and the two shaded intervals are the
time windows $W_1$ and $W_2$ used for the diffusion-coefficient
inversion. The grey band in (a) marks the inter-shell radiative layer,
as in Fig.~\ref{fig:profiles}.}
\label{fig:velocity}
\end{figure*}

The 3D simulations use a radius-only fit of the \mesa{} opacity, of
functional form
\begin{equation}
\label{eq:kappa_fit}
\begin{aligned}
\kappa(R) ={}& a\,R \;+\; b\,(R-c)\,\tanh\!\bigl[d\,(R-c)\bigr] \\
             & {}+\; e\,\tanh\!\bigl[f\,(R-g)\bigr] \;+\; h,
\end{aligned}
\end{equation}
with eight fit parameters $a,b,c,d,e,f,g,h$
(Fig.~\ref{fig:setup_inputs}a). In the 1D base state the convectively
unstable C-burning region extends out to ${\sim}65\usp\Mm$ as an
essentially isentropic layer (Fig.~\ref{fig:base_state}a). Including
the full C-shell at the grid spacings used here would leave too few
cells across the inter-shell radiative layer and the $N^2$ peak to
resolve them,
so both envelope configurations truncate it well short of its full
radial extent, while the inter-shell radiative layer and its
IGW field ($r\approx 9\text{--}10\usp\Mm$) are fully contained in
either setup.
The base state loaded into the 3D runs
retains the $N^2$ peak of the 1D \mesa{} model, though the setup's
spline smoothing lowers and broadens it by roughly a factor of two in
$N$ (Fig.~\ref{fig:base_state}b, with the same base-state smoothing
described in the appendix of \citealt{Herwig:23a}). The time-averaged
Mach-number profiles (Fig.~\ref{fig:velocity}a) show the two driven
shells as high-Mach plateaus ($\mach\approx 6\times10^{-3}$ in the
O-shell) on either side of the inter-shell radiative layer, where the
Mach number drops by about an order of magnitude. The no-heating control
O31 collapses to far lower velocities. 
The burning run O43 follows the
heated runs.

We implement the $^{12}$C+$^{12}$C burn following the approach
of \citet{Andrassy:2020} for their NET 2, without $^{12}$C+$^{16}$O. This network uses an effective $Q$-value of
$3.19\usp\mathrm{MeV}$ that is the sum over the (n, p, $\gamma$)
$Q$-values from \citet{CF88},
\begin{equation}
\label{eq:eps_nuc}
\epsilon_{\mathrm{nuc}} = b\,\frac{1}{2}\,Q_{12}\,N_{\mathrm{A}}\,\rho\,
Y(^{12}\mathrm{C})^{2}\,\lambda_{12},
\end{equation}
where $b$ is a boost factor, $Q_{12}$ is the effective $Q$-value,
$N_{\mathrm{A}}$ is Avogadro's number, $\rho$ is the density,
$Y(^{12}\mathrm{C})$ is the molar abundance of $^{12}$C, and
$\lambda_{12}$ is the effective rate. The nuclear burn is restricted to
$r \geq 9.42\usp\Mm$ and scaled by $b = 2.6$ to reproduce
the \mesa{} $\epsilon_\mathrm{nuc}$ profile at the shell
(Fig.~\ref{fig:setup_inputs}b).

\FloatBarrier
\section{Supplementary IGW Diagnostics}
\label{app:wave}

\begin{figure*}[tb]
\centering
\includegraphics[width=\textwidth]{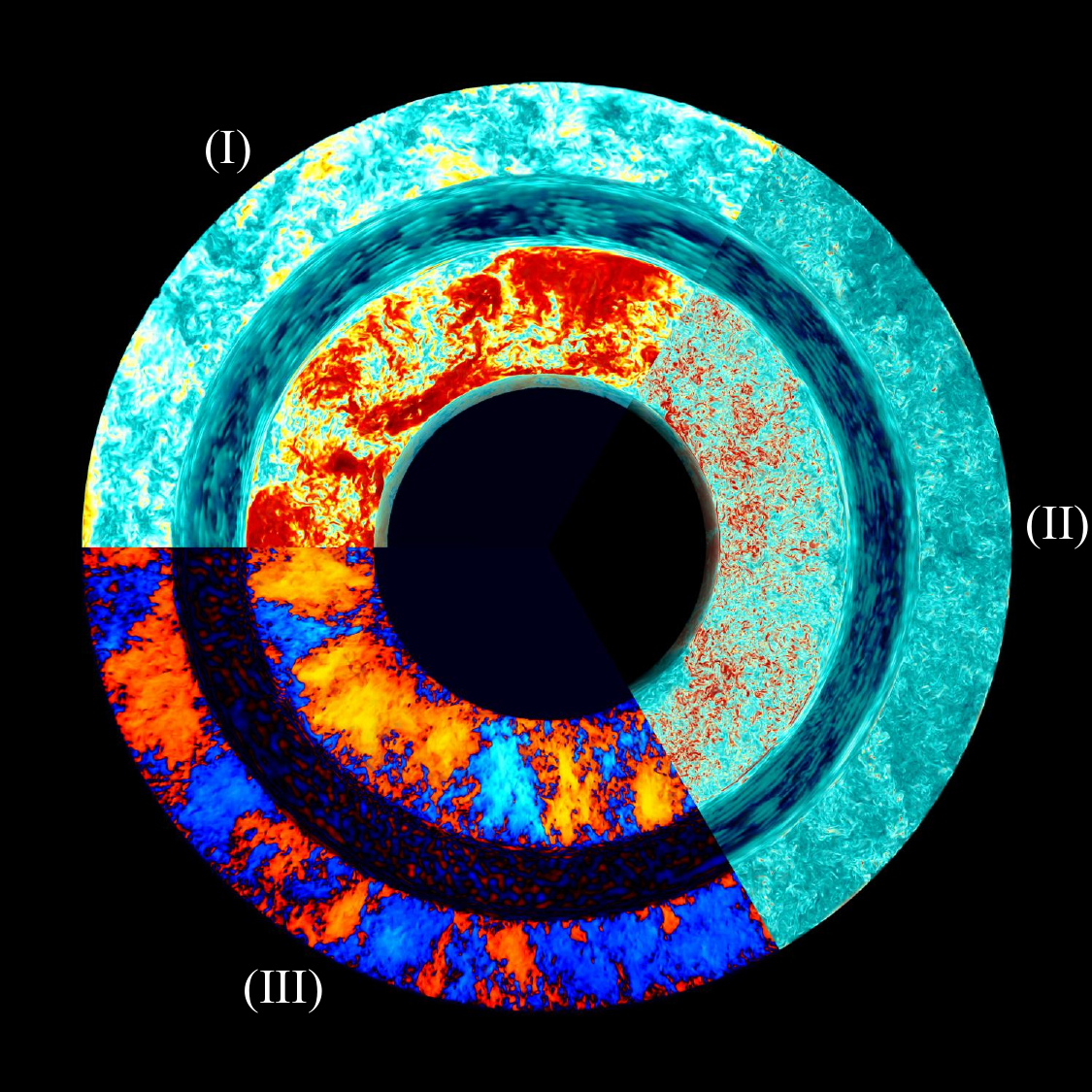}
\caption{Volume renderings of three fluid variables in a thin
equatorial slice of run O32 ($3072^3$) at $t=0.86\usp\hour$. (I)
Horizontal velocity magnitude, colored from highest to lowest in dark
brown, red, yellow, white, light blue and dark blue. (II) Vorticity
magnitude, colored from highest to lowest in red, yellow, light blue and
dark blue. (III) Radial velocity, with inward (negative) velocities in
light to dark blue and outward (positive) velocities in yellow, orange
and red, each in order of decreasing magnitude. The inner O-shell and outer C-shell appear as
turbulent rings on either side of the large-scale IGW field of
the inter-shell radiative layer. Figure~\ref{fig:vort_zoom} is a close-up of wedge (II).}
\label{fig:render_O32}
\end{figure*}

\begin{figure*}
\centering
\includegraphics[width=\linewidth]{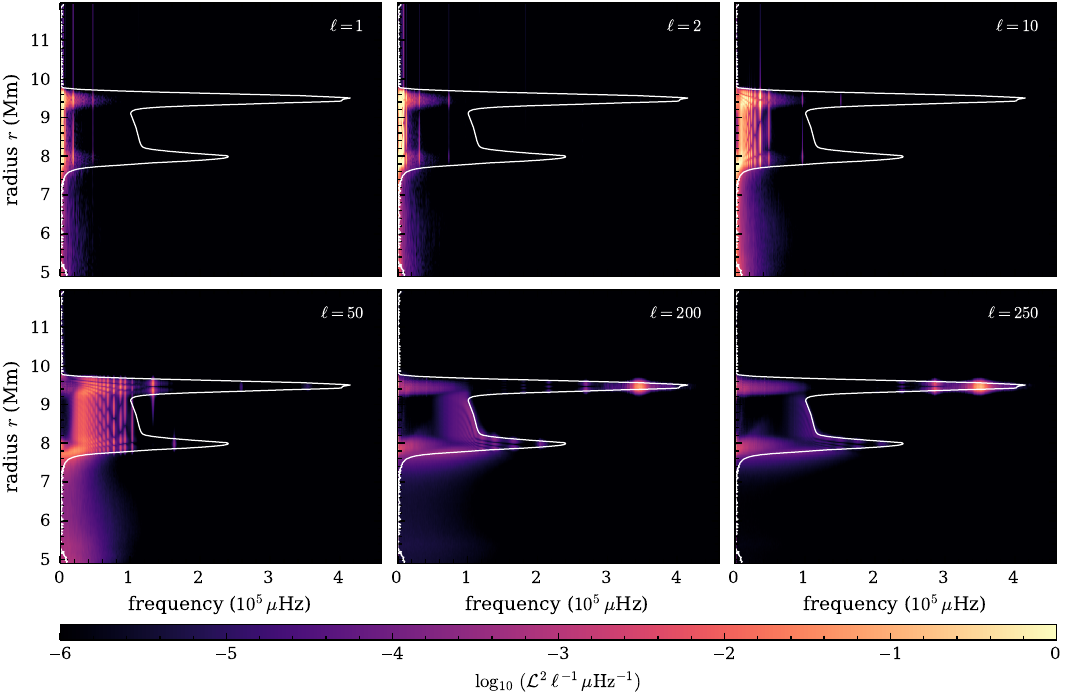}
\caption{Wave-propagation (radius versus frequency) diagrams of the
unity-subtracted relative luminosity $\mathcal{L}$ in O28 for six
angular degrees $\ell$ (panel labels). Within each panel the
horizontal axis is frequency ($10^{5}\usp\mathrm{\mu Hz}$) and the
vertical axis is radius (Mm). The white curve is the \brunt{} frequency
profile $N(r)$.}
\label{fig:wave_prop}
\end{figure*}

Equatorial-slice renderings of the horizontal velocity, vorticity
and radial velocity from O32 (Fig.~\ref{fig:render_O32}) show the inner
O-shell and outer C-shell as turbulent rings, between which the radiative
layer carries a large-scale concentric wave pattern. Figure~\ref{fig:vort_zoom} is a close-up
of the vorticity wedge. A radius--frequency propagation
diagram for selected angular degrees (Fig.~\ref{fig:wave_prop}) shows
the structure of the layer.
Convective power dominates the O-shell ($R<7.8\usp\Mm$), and
discrete resonant eigenmodes fill the radiative layer
($7.8\lesssim R\lesssim 10\usp\Mm$) below the \brunt{} frequency and the
Lamb frequency $S_\ell$. In the layer $S_\ell$ exceeds
${\sim}5\times10^{5}\usp\mathrm{\mu Hz}$ even at $\ell=1$ and rises
steeply with $\ell$. Toward high $\ell$ the modes concentrate near the
\brunt{} maximum, consistent with the $\ell$--$\nu$ diagrams.
Inside the C-shell convection zone the spectrum is dominated by
low-frequency convective power, with only evanescent IGW and
p-mode-like features, as in the outer convection zones of full-star
\ppmstar{} models \citep{Pathak:2025}.

\FloatBarrier
\section{The Heating/No-Heating Control}
\label{app:control}

\begin{figure*}[tb]
\centering
\includegraphics[width=\linewidth]{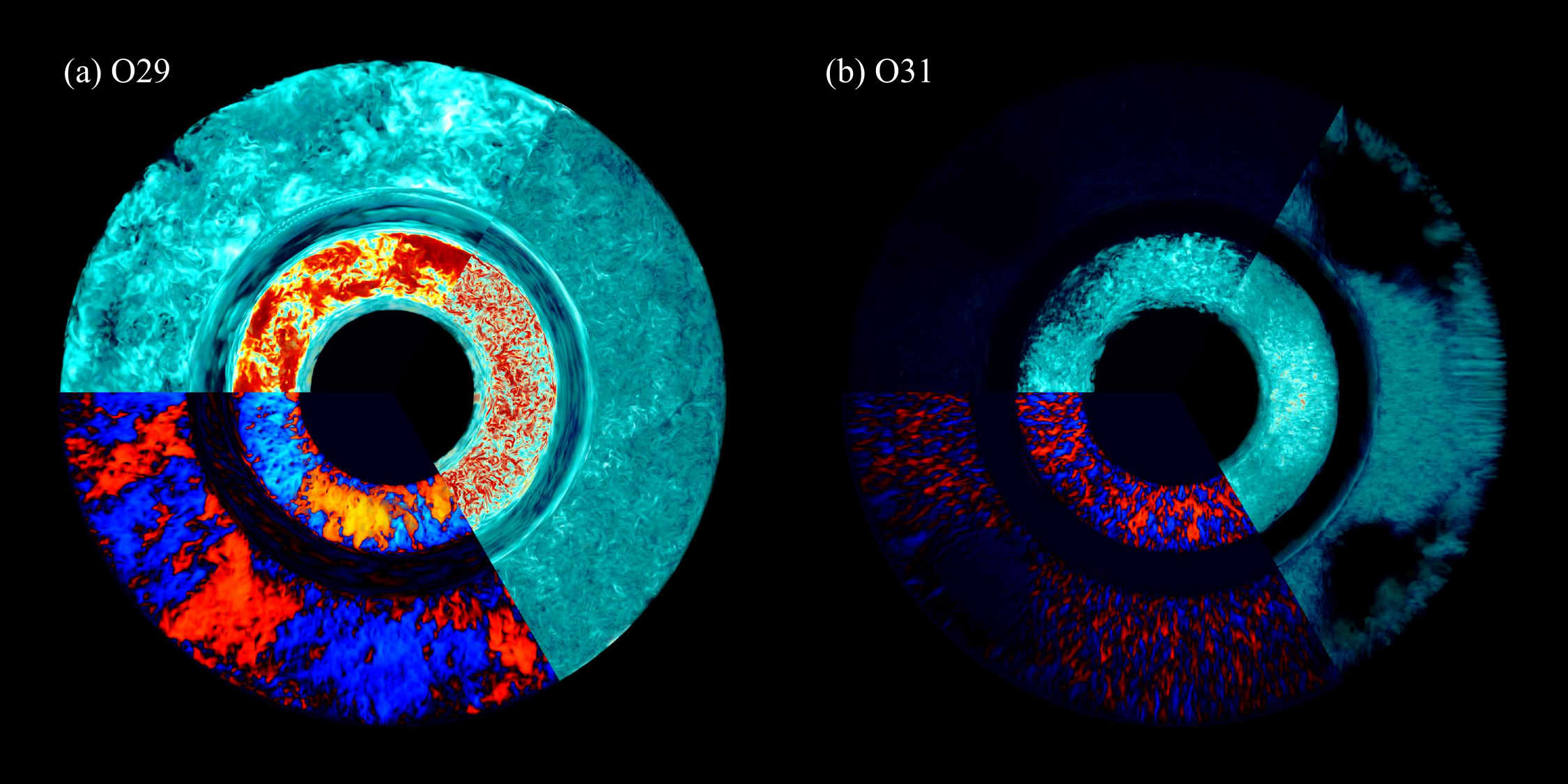}
\caption{Volume renderings of run O29 (a, heated) and its
no-heating control O31 (b) at $t\approx1.67\usp\hour$, in the same three
fluid-variable wedges as Fig.~\ref{fig:render_O32}. Both runs use a
$1536^3$ grid and the large envelope ($R_\mathrm{max}=16\usp\Mm$).}
\label{fig:render_control}
\end{figure*}

Figure~\ref{fig:render_control} contrasts run O29 (heated) with its
no-heating control O31 at $t\approx1.67\usp\hour$, using the same three
fluid-variable wedges as Fig.~\ref{fig:render_O32}. O29 sustains
vigorous convection and a turbulent inter-shell IGW field. The
no-heating control O31 is nearly still, with the convective driving and
the inter-shell motions collapsed.

\begin{figure*}[tb]
\centering
\includegraphics[width=\textwidth]{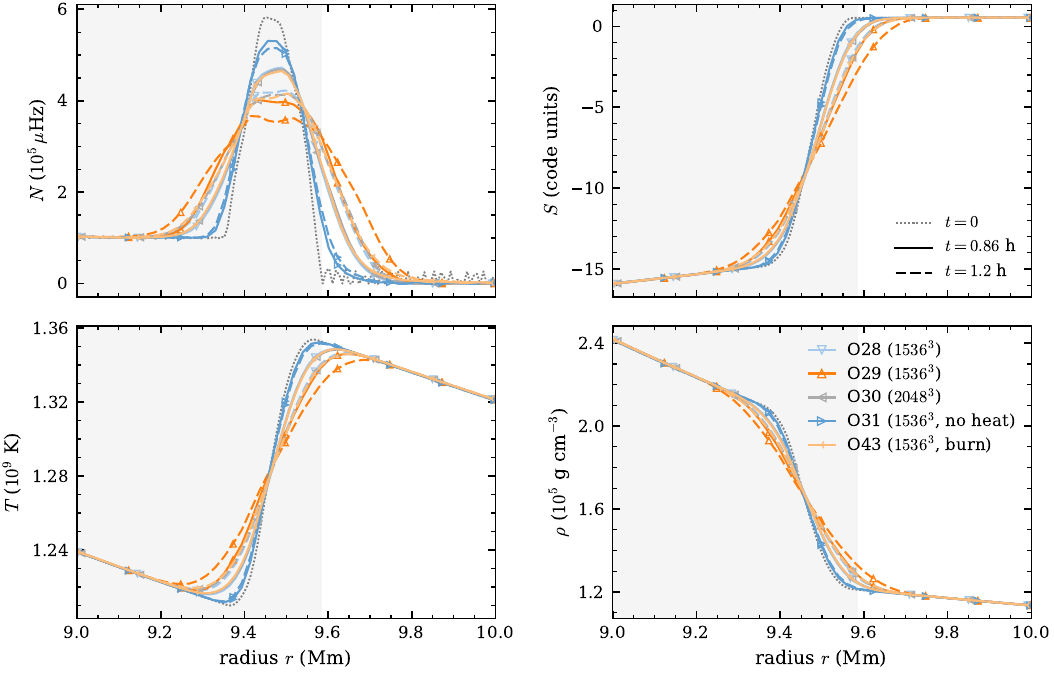}
\caption{Radial profiles around the $N^2$ peak for
the large-envelope runs O29, O30 and the no-heating control O31, with
the small-envelope O28 and the active C-burning run O43 for comparison,
in the layout and line styles of Fig.~\ref{fig:profiles}.}
\label{fig:profiles_env}
\end{figure*}

Figure~\ref{fig:profiles_env} shows the inter-shell profiles of the
large-envelope runs O29, O30 and O31, with the small-envelope O28 and the
burning run O43 for comparison, at the same two times and with the same
layout as Fig.~\ref{fig:profiles}. The heated large-envelope runs erode
the $N$ peak like the small-envelope sequence. O29 and O30 fall to
$4.0$ and $4.7\times10^{5}\usp\mathrm{\mu Hz}$ by $0.86\usp\hour$ and to
$3.7$ and $4.1\times10^{5}$ by $1.2\usp\hour$, at or below the O28 values ($4.73$ and
$4.22\times10^{5}$). The
no-heating control O31 stays close to its $t=0$ state, at $5.3$ and
$5.2\times10^{5}\usp\mathrm{\mu Hz}$ at the same times, and its $T$, $S$
and $\rho$ steps remain sharp.

\FloatBarrier
\section{Diffusion-Coefficient Method and Window Dependence}
\label{app:dmethod}

\begin{figure*}
\centering
\includegraphics[width=\linewidth]{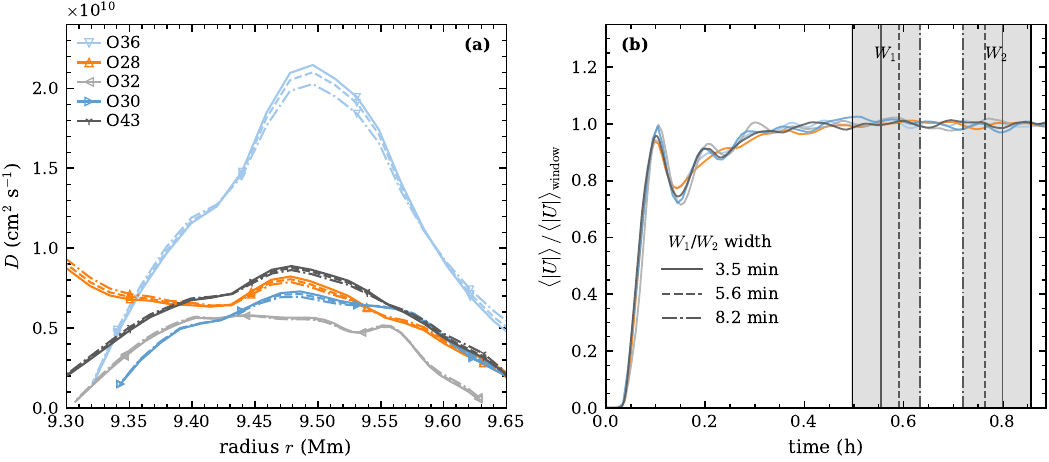}
\caption{Dependence of the entropy-mixing diffusion coefficient $D(r)$
on the temporal averaging-window width. (a) $D(r)$ over
$R\in[9.30,9.65]\usp\Mm$ for O36, O28, O32, O30 and O43 (color and
marker), each
computed with three equal widths for the early ($W_1$) and late
($W_2$) averaging windows (line style). (b) The normalized
convective velocity $\langle|U|\rangle$ time series of each run, with
the $W_1$ and $W_2$ windows for the three widths shaded. All runs use the common time interval of
Fig.~\ref{fig:Dcombined}b.}
\label{fig:Dwindow}
\end{figure*}

The early ($W_1$) and late ($W_2$) inversion windows are taken
from the plateau that the volume-averaged O-shell velocity reaches after
the initial transient (Fig.~\ref{fig:velocity}b). The inversion of \S\ref{sec:Dcoeff} averages the entropy
variable $S$ over $W_1$ and $W_2$ to give $\bar S_1$ and
$\bar S_2$, and infers $D(r)$ from the difference of the two profiles.
The common width of $W_1$ and $W_2$ is a free parameter of
the method (described in detail by \citealt{Herwig:23a}). To test whether the inferred coefficient depends on this
width, we repeat the inversion for all five runs (the small-envelope resolution series O36,
O28 and O32, the large-envelope run O30 and the burning run O43) over the same common time
interval used in Fig.~\ref{fig:Dcombined}b (set by the shortest run,
O32), for three equal $W_1$/$W_2$ widths spanning a factor of
${\sim}2.4$ (${\approx}3.5\text{--}8.2\usp\minute$ of simulated time, or
${\approx}16\text{--}38\%$ of the usable interval). $W_1$ is anchored at
the start and $W_2$ at the end of the interval, and the two are kept
non-overlapping, so the inversion baseline $t_2-t_1$ stays well defined
(it shrinks from ${\approx}18$ to ${\approx}13\usp\minute$ as the windows
widen; Fig.~\ref{fig:Dwindow}b). Figure~\ref{fig:Dwindow}a shows the
resulting $D(r)$. Within each run the curves for the different widths
are very similar. The peak value changes by at most a few percent, and
by ${<}1\%$ for the adopted highest-resolution run O32, while the
run-to-run differences in resolution and envelope are unchanged. The
inferred entropy-mixing coefficient at the $N^2$ peak is therefore converged
with respect to the averaging-window width.

\begin{figure*}[tb]
\centering
\includegraphics[width=\textwidth]{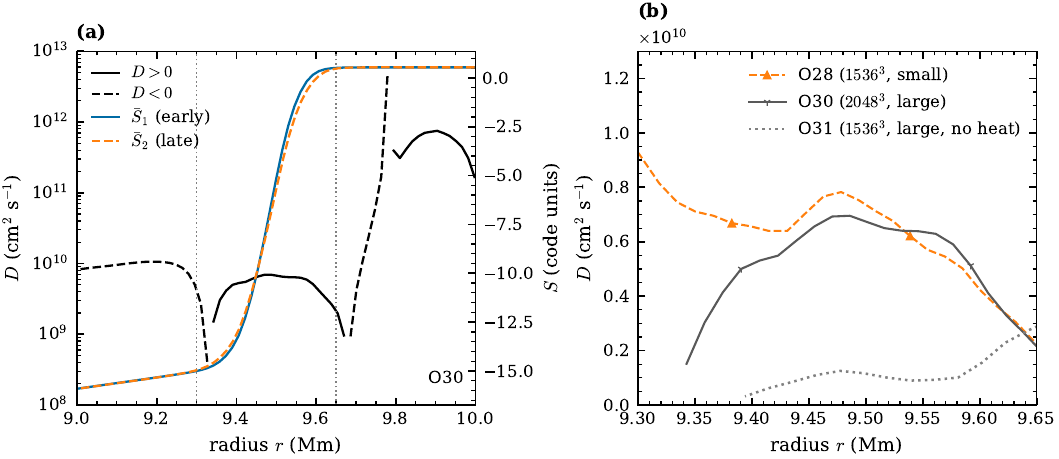}
\caption{Envelope independence of the mixing at the $N^2$ peak. (a) Inversion
for the large-envelope run O30 ($2048^3$, $R_\mathrm{max}=16\usp\Mm$),
in the layout of Fig.~\ref{fig:Dcombined}a. (b) $D(r)$ over the
window for O30 and the small-envelope O28 ($1536^3$,
$R_\mathrm{max}=12\usp\Mm$), which share the grid spacing
$\Delta x = 0.0156\usp\Mm$, and for the no-heating control O31
(dotted).}
\label{fig:D_envelope}
\end{figure*}

Figure~\ref{fig:D_envelope} tests whether the envelope size affects the
inferred mixing. O28 and O30 share the grid spacing
$\Delta x = 0.0156\usp\Mm$ but extend to $R_\mathrm{max}=12$ and
$16\usp\Mm$, respectively. Inverted over the same common interval and
with the same early and late windows as Fig.~\ref{fig:Dcombined}, they
give peak $D$ of $7.8\times10^{9}$ and
$7.0\times10^{9}\usp\power{\cm}{2}\usp\power{\second}{-1}$, an 11\%
difference, with the same radial shape across the window
(Fig.~\ref{fig:D_envelope}b). The additional $4\usp\Mm$ of C-shell
envelope therefore does not change the mixing at the $N^2$ peak
significantly. Inverted in the same way, the no-heating control O31 gives a
small $D(r)$ that changes sign across the window, with a peak of
$1.2\times10^{9}\usp\power{\cm}{2}\usp\power{\second}{-1}$, 16\% of
O28's, consistent with its near-static profiles
(Appendix~\ref{app:control}, Fig.~\ref{fig:profiles_env}).

\FloatBarrier
\section{A Direct Diffusivity Estimate from the Inert-Tracer (FV) Peaks}
\label{app:fv}

\begin{figure*}
\centering
\includegraphics[width=\linewidth]{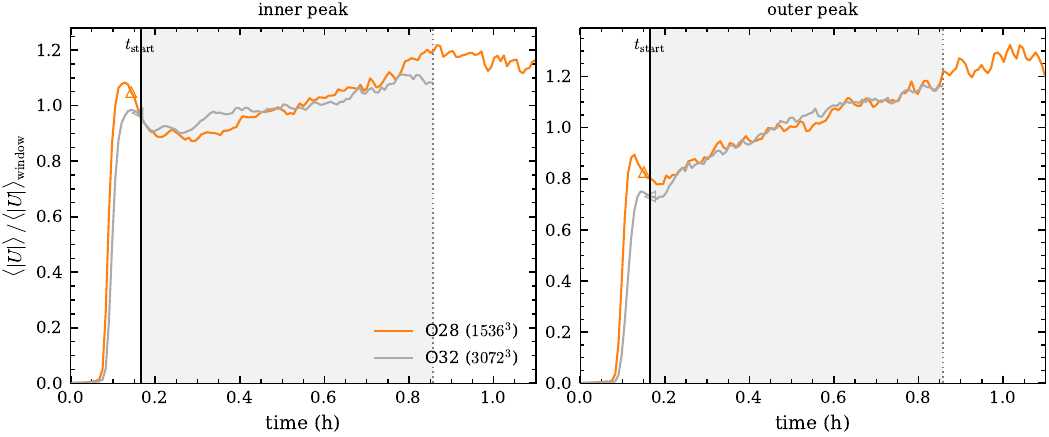}
\caption{The velocity transient that sets $t_\mathrm{start}$. For each
FV Gaussian peak (inner, outer panels), the spatial
average of $|U|$ over the peak's FWHM region, normalized to its mean
over the fit window, versus time, for O28 ($1536^3$) and O32 ($3072^3$).
$\langle|U|\rangle$ rises rapidly from rest and overshoots. The open
marker on each curve marks the end of that initial transient (its first
turnover), and the solid vertical line is the adopted $t_\mathrm{start}$.
The shaded band is the fit window $[t_\mathrm{start},t_\mathrm{end}]$
and the dotted line is $t_\mathrm{end}$.}
\label{fig:fvhhe_u}
\end{figure*}

Two Gaussian profiles of an inert
tracer fluid (FV), advected with the high-order piecewise-parabolic
Boltzmann (PPB) scheme \citep{Woodward:15}, are placed inside the
radiative layer to measure $D$ there directly, where the $S$-variable
diffusion-equation inversion of \S\ref{sec:Dcoeff} is less accurate. The
two FV peaks are centered at $r\approx8.2$ and $9.1\usp\Mm$. This method
of estimating $D$ from the erosion of tracer peaks is also used by
\citet{Blouin_RGB:25}. Treating each peak as a Gaussian that
broadens diffusively ($\sigma^2(t)=\sigma_0^2+2Dt$, so that the
area-conserving amplitude declines as $A\propto\sigma_0/\sigma$), the
diffusion coefficient at the peak follows from the drop of its maximum
value,
\begin{equation}
\label{eq:fvhhe_D}
D = -\,\sigma_0^2\,\frac{1}{FV_{\max,0}}\,\frac{\mathrm{d}FV_{\max}}{\mathrm{d}t},
\end{equation}
where $\sigma_0$ is the standard deviation of the fitted Gaussian at the start of the
fit and the slope is taken over a window
$[t_\mathrm{start},t_\mathrm{end}]$. We fix $t_\mathrm{start}$ from the
velocity field. For each peak we average $\langle|U|\rangle$ over its FWHM
region (Fig.~\ref{fig:fvhhe_u}) and identify the end of the initial
transient as the first
turnover of the rapidly rising $\langle|U|\rangle$. We then take
$t_\mathrm{start}$ as the maximum of these over both peaks and both
runs (${\approx}0.17\usp\hour$),
and $t_\mathrm{end}$ is the end of the shortest run O32, as in
Fig.~\ref{fig:Dcombined}b. We restrict the analysis to the two
highest-resolution small-envelope runs, O28 ($1536^3$) and O32 ($3072^3$).

\begin{figure*}
\centering
\includegraphics[width=\linewidth]{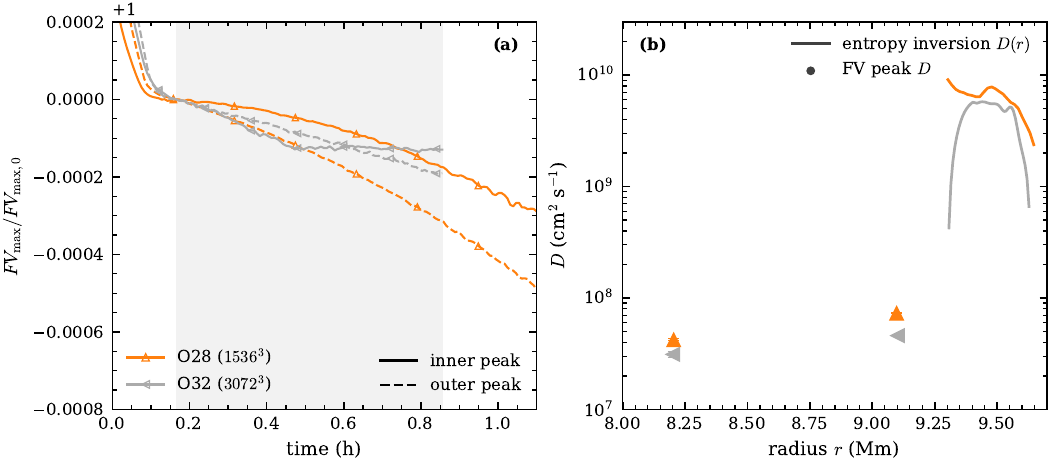}
\caption{Direct diffusivity at the inert-tracer (FV) peaks for
O28 ($1536^3$) and O32 ($3072^3$). (a) The peak amplitude
$FV_{\max}/FV_{\max,0}$, normalized at $t_\mathrm{start}$, versus time
for the inner (solid) and outer (dashed) peaks (note the offset,
expanded ordinate). The shaded band is the analyzed window
$[t_\mathrm{start},t_\mathrm{end}]$ over which the slope, and hence
$D$ (Eq.~\ref{eq:fvhhe_D}), is fitted. (b) The resulting $D$ at the two FV Gaussian
peaks ($r\approx8.2$ and $9.1\usp\Mm$; markers, with error
bars from the slope fit) compared with the entropy-inversion
$D(r)$ at the $N^2$ peak of Fig.~\ref{fig:Dcombined}b over $R\in[9.30,9.65]\usp\Mm$ (lines).}
\label{fig:fvhhe}
\end{figure*}

Over the fit window the peak amplitudes decline by only a few parts in
$10^{4}$ (Fig.~\ref{fig:fvhhe}a).
Figure~\ref{fig:fvhhe}b shows the resulting peak-$D$ values together
with the entropy-inversion $D(r)$ of Fig.~\ref{fig:Dcombined}b. The direct
estimate, $D\approx(3\text{--}7)\times10^{7}\usp\power{\cm}{2}\usp\power{\second}{-1}$ at
the two peak radii, is small and decreases with resolution (O32 below O28
at both peaks), mirroring the downward convergence of the $S$-based
coefficient, though it sits well below the $S$-inversion value at the $N^2$ peak.
The tracer is carried by the higher-order PPB advection while the
thermodynamic variables are carried by the more diffusive PPM scheme, so
this estimate is a lower bound on the numerical diffusion acting on the
entropy. The
FV peaks remain saturated near their initial value (${\approx}0.9$), so
their \emph{amplitude} barely changes while the erosion is confined to
the flanks. The FV peaks nonetheless indicate that the interior of the
radiative layer, away from the $N^2$ peak, is weakly mixed, with the
peaks remaining essentially stationary over the analyzed interval.

\FloatBarrier
\section{Forward Extrapolation of the Entropy Profile}
\label{app:forward}

The extrapolation of \S\ref{sec:fate} evolves the spherically
averaged entropy variable $S$ with the diffusion equation
\begin{equation}
\label{eq:forward}
\frac{\partial S}{\partial t} = \frac{1}{r^{2}}\,\frac{\partial}{\partial r}
\left( r^{2} D(r,t)\,\frac{\partial S}{\partial r} \right),
\end{equation}
started from the late-window average $\bar S_2(r)$ of the inversion at
$t_2=0.79\usp\hour$. Equation~\ref{eq:forward} is discretized in
conservative form on a uniform grid of spacing $0.002\usp\Mm$, with the
harmonic mean of $D$ at the cell interfaces, and integrated with the
Crank--Nicolson scheme at a time step of $5\usp\second$, the operator
rebuilt at the start and at the end of every step. O28 is used because
its $D(r)$ is close to the converged O32 profile
(Fig.~\ref{fig:Dcombined}b) and because it runs to $1.76\usp\hour$,
twice as long as O32, which leaves a stretch of simulated evolution
against which the model can be tested before it is extrapolated.

\begin{figure*}
\centering
\includegraphics[width=\textwidth]{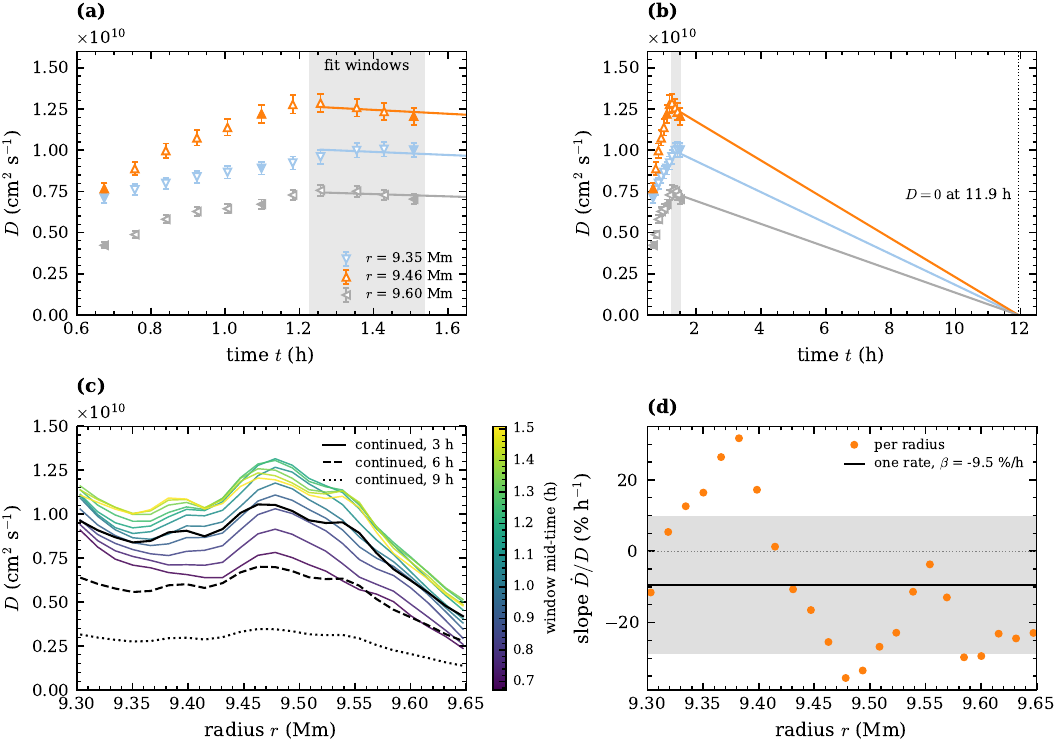}
\caption{The time dependence of $D$ in O28 and its
extrapolation. (a) $D$ at three radii against the mid-time of the
window pair, for the eleven pairs of the sliding inversion. The four
fit windows are shaded, the three pairs sharing no dumps are filled,
and the error bars are the $4\%$ spread across window widths of
Fig.~\ref{fig:Dwindow}. Lines are Eq.~\ref{eq:decline}. (b) The same
lines continued to $D=0$ at $11.9\usp\hour$. (c) The measured $D(r)$ at
the eleven times (color) and the continued profile at $3$, $6$ and
$9\usp\hour$. (d) The relative slope of $D$ over the fit windows at
each radius, the single rate $\beta$ (line) and the scatter of the
per-radius values (band).}
\label{fig:D_extrap}
\end{figure*}

The inversion of
\S\ref{sec:Dcoeff} uses the interval common to all runs, which the
length of O32 sets. O28 itself runs almost an hour longer, and its
entropy gradient keeps flattening, so we measure how $D$ responds by
sliding the two inversion windows of Fig.~\ref{fig:Dcombined}b, $8$
minutes wide and $13.5$ minutes apart, forward through the run in
$5$-minute steps. This gives $D(r)$ at eleven times between $0.67$ and
$1.51\usp\hour$ (Fig.~\ref{fig:D_extrap}), the first pair being the
inversion of Fig.~\ref{fig:Dcombined}b itself. Three of the eleven
share no dumps. As the gradient flattens $D$ rises at every radius, by
a factor $1.7$ at the peak and $1.6$ in its integral over the window,
up to $1.2\usp\hour$, and declines after. The wave velocity amplitude
in the layer rises by $27\%$ over the same interval.
The extrapolation uses only the run's own late trend. Over the four windows after the peak, at $1.26$--$1.51\usp\hour$,
one relative rate fitted to the whole profile by least squares over all
radii and windows,
\begin{equation}
\label{eq:decline}
D(r,t) = \bar D(r)\,\bigl[1+\beta\,(t-\bar t)\bigr],\qquad
\beta = -0.095\usp\power{\hour}{-1},
\end{equation}
with $\bar D(r)$ the mean profile of those windows and $\bar t$ their
mean time, describes them to $2\%$. Before the fit windows the measured
profiles are interpolated in time. After them Eq.~\ref{eq:decline} is
followed until $D$ reaches zero at $11.9\usp\hour$, and $D$ stays zero.
Fitted radius by radius, the slopes scatter by $\pm 20\%$ per hour
about $\beta$, positive inside $9.40\usp\Mm$ and negative outside
(Fig.~\ref{fig:D_extrap}d). Continued independently they cross zero at
different times and leave holes in $D(r)$, so the one rate is applied
to the profile with its shape held. Only the radii where every window
has $D>0$, $9.303$--$9.647\usp\Mm$, enter. Outside them the edge value
is held.

Over the whole run the entropy at $9.2$ and
$9.8\usp\Mm$ changes by $0.02$ units, while at the window edges,
$9.30$ and $9.65\usp\Mm$, it drifts by $0.4$ and $0.7$ units as the
mixed entropy spreads into the wings. Figure~\ref{fig:Dcombined}c passes
no entropy flux through the window edges.
Figure~\ref{fig:Dcombined}d holds $S$ at its $t_2$ values at $9.2$ and
$9.8\usp\Mm$. At $1.76\usp\hour$ both
reproduce the simulated profile over the window to an rms of $0.75$
and $0.64$ units of the $15.4$-unit step, against $0.68$ for constant
$D$, but the entropy at the window edges separates them. The simulation
gives $-14.54$ and $-0.25$ at $9.303$ and $9.647\usp\Mm$, the held
reservoirs $-14.34$ and $-0.20$, and the zero-flux window $-13.47$ and
$-0.85$, entropy piling up at edges it cannot cross.

By the time $D$ reaches zero the step across
the window has shrunk from $15.4$ to $2.6$ units with zero-flux edges
and to $9.6$ units with the reservoirs held, and no longer changes.
With $D(r)$ held constant at its inverted value instead, the zero-flux
window flattens completely, $0.2$ units at $24\usp\hour$, while the
held reservoirs leave a linear ramp of $7.5$ units across the window at
$48\usp\hour$, the flux-conserving steady state between fixed end
values. The edge treatments bound the response of the two convection
zones, which the model leaves out. The $25\%$ lower converged $D$ of
O32 would leave correspondingly more of the step.

\bibliography{hydropaper-resources/bib/hydro,oshell_local}
\bibliographystyle{aasjournalv7}

\end{document}